\documentclass[letterpaper]{aa}
\usepackage{graphicx}

\begin{document}
   \title{The young active binary star \object{EK~Draconis}$^*$}

   \subtitle{}

   \author{B. K\"onig\inst{1,2}
          \and
           E.W. Guenther\inst{3}
          \and
           J. Woitas\inst{3}
          \and
           A.P. Hatzes\inst{3}
             }

   \offprints{Brigitte K\"onig,
     \email{bkoenig@bruno.phyast.pitt.edu}\\ $*$~Partly based on
     observations collected at the German-Spanish Astronomical Center,
     Calar Alto, operated jointly by the Max-Planck Institut f\"ur
     Astronomie and the Instituto de Astrof\'{\i}sica de
     Andaluc\'{\i}a (CSIC), and partly based on observations taken
     with the 2-m-Alfred Jensch telescope of the Th\"uringer
     Landessternwarte Tautenburg.}

   \institute{Max-Planck-Institut f\"ur extraterrestrische Physik,
              Gie\ss enbachstra\ss e 1, D-85748 Garching, Germany
                  \and
              University of Pittsburgh, 3941 O'Hara St, Pittsburgh, PA 15260, 
              USA  
                  \and
              Th\"uringer Landessternwarte Tautenburg, 
              Sternwarte 5, D-07778 Tautenburg, Germany 
                           }

   \date{Received; accepted}

   \abstract{\object{EK Dra} (\object{HD\,129333}) is a young, active,
    nearby star that is orbited by a low mass companion. By combining
    new speckle observations with old and new radial velocity
    measurements we find that the orbit is highly eccentric with
    $e=0.82\pm0.03$, and we derive the true masses of both
    components. The masses are $0.9\pm0.1\,{\rm M}_\odot$ and
    $0.5\pm0.1\,{\rm M}_\odot$, for the primary and secondary,
    respectively. From high resolution spectra we derive a new $T_{\rm
    eff}$ of $5700\pm70$~K and a $\log g$ of $4.37\pm0.10$, which is
    different to previous estimates. However, the new spectroscopic
    distance differs by only 5.8\% to the distance derived by parallax
    measurement by the {\it Hipparcos} satellite and thus the stellar
    parameters are presumably more realistic than older
    determinations. We derive a somewhat higher value for the
    metallicity of $[Fe/H]=-0.16\pm0.07$. \object{EK Dra} turns out to
    be one of the few nearby young stars that will evolve similarly to
    the Sun. The precise radial velocity measurements taken in the
    course of this program also allow us to shed more light on the
    activity of this star. In 2001 and 2002 we find radial velocity
    variation with a period of $2.767\pm0.005$\,days which we
    interpret as the rotation period. This signal vanishes in
    2003. However the signal can be recovered if only the spectra in
    which the photospheric lines are asymmetric are used. On the other
    hand, we do not find a close correlation between the asymmetry of
    photospheric lines and the radial velocity.  } \maketitle

%

\section{Introduction}
\object{EK Dra} (GJ 559.1A, HD 129333) is a star that has roughly the
mass of the Sun. The Henry Draper Catalogue and Extension
(\cite{Cannon1924}) and Simbad database list a spectral type of F8 for
this star. The equivalent width of Li\,I is about 0.2\,\AA, and
therefore it must be relatively young. Fr\"ohlich et
al. (\cite{froehlich02}) and references therein discuss the age and
activity connection.

Various studies of \object{EK~Dra} at different wavelength regimes have been
carried out including longterm photometric monitoring over decades. Dorren \&
Guinan (\cite{dorren94}) have observed strong variable chromospheric emission
lines in their UV spectra. The star has the highest known Ca\,II H and K
emission level of any known early G-type star which is not a close binary
(Soderblom \cite{soderblom85}).

The star is rapidly rotating ($v \sin i =16.5\pm1.0$\,km/s) and has
dominant spot features at $\sim 70^\circ - 80^\circ$\ that could also
be the offshoot of a large polar spot. These spots are located at a
higher latitude than typical spots on the Sun (Strassmeier \& Rice
\cite{strassmeier98}). Strassmeier \& Rice (\cite{strassmeier98})
  measure several rotation periods between $2.599\pm0.001$\,days and
  $2.796\pm0.026$\,days using different methods where for their
  purpose they adopt a longterm photometric period of 2.605\,days.
Coronal emission was also observed in X-rays and as well in the radio
regime. The X-ray light curve is significantly variable, with the
emission from the cooler plasma being strongly modulated by the
rotation period, while the emission from the hotter plasma is only
weakly variable (Guedel et al.  \cite{guedel95}).

A 12 to 14 year cyclic variability was discovered by Dorren \& Guinan
(\cite{dorren94}) and Dorren et al. (\cite{dorren95}) using photometric and
spectroscopic data. They observed that the Ca\,II~H \&~K emission index
increased during that time.  A decline of brightness since 1994 was noted by
Fr\"ohlich el al. (\cite{froehlich02}). The star became fainter as its mean
level of chromospheric activity rose. These findings can be interpreted as
signs of a spot cycle.

\object{EK Dra} also is a long period binary star where the secondary is much
fainter than the primary.  Duquennoy \& Mayor (\cite{duquennoy91}) used a
period of 11.5 years in order to derive the first preliminary spectroscopic
orbit. However, as will be discussed in the next section, the true period
is $45\pm5$ years.

\object{EK Dra} is a well-studied young, active and nearby star. It
thus serves as one of the best-studied young stars evolving similar to
the Sun. Since it is a long period binary, it is possibly one of the
few cases of young stars for which the true masses can be
determined. The aim of this paper is to derive the true mass, and to
calculate a new atmospheric model. These will then allow us to compare
the properties of this object with evolutionary tracks. Additionally,
we have obtained a large number of radial velocity measurements which
will give us new insights into the stellar activity and the influence of
stellar activity on precise radial velocity measurements.


\section{Deriving the true masses of EK\,Dra\,A and  EK\,Dra\,B}

By combining the data from our speckle interferometry and RV data from the
literature with our own RV measurements, it is for the first time possible to
derive the true masses of EK\,Dra\,A and EK\,Dra\,B.

\subsection{Speckle interferometry : observations}
\begin{table*}
\caption{An overview of all spatially resolved observations
 of the binary system \object{EK Dra }AB, carried out with near-infrared speckle
 interferometry at the 3.5-m telescope on Calar Alto.}
\begin{tabular}{llllllll}
\hline \hline 
No. & Date & Epoch & Position & Projected & Filter & Flux ratio & Instrument \\
    &      &       & angle [$^{\circ}$] & separation [mas] & & $F_B/F_A$ & \\
\hline 
1 & 19.03.1991 & 1991.2135 & 173.9 $\pm$ 4.3 & 282 $\pm$ 12 & K & & 1D \\
2 & 13.02.1992 & 1992.1232 & 171.7 $\pm$ 3.9 & 313 $\pm$ 13 & K & & 1D \\
3 & 05.10.1993 & 1993.7611 & 173.7 $\pm$ 0.9 & 456 $\pm$ 14 & K &
    0.066 $\pm$ 0.003 & MAGIC \\
4 & 26.01.1994 & 1994.0712 & 174.0 $\pm$ 0.3 & 486 $\pm$ 10 & K &
    0.084 $\pm$ 0.003 & MAGIC \\
5 & 13.12.1994 & 1994.9501 & 175.5 $\pm$ 0.4 & 501 $\pm$ 5  & H &
    0.087 $\pm$ 0.003 & MAGIC \\
6 & 22.11.1997 & 1997.8926 & 174.8 $\pm$ 0.6 & 647 $\pm$ 12 & K &
    0.084 $\pm$ 0.004 & MAGIC \\
7 & 10.02.2001 & 2001.1123 & 172.8 $\pm$ 0.6 & 679 $\pm$ 5  & K &
    0.111 $\pm$ 0.006 & OMEGA Cass \\
8 & 03.11.2001 & 2001.8406 & 174.0 $\pm$ 1.0 & 709 $\pm$ 10 & K &
    0.081 $\pm$ 0.008 & OMEGA Cass \\
9 & 21.10.2002 & 2002.8049 & 171.9 $\pm$ 1.0 & 725 $\pm$ 8  & K &
    0.081 $\pm$ 0.010 & OMEGA Cass \\
\hline\hline
\end{tabular}
\label{speckle_obs} 
\end{table*}

\begin{figure}[h]
\resizebox{\hsize}{!}{\includegraphics{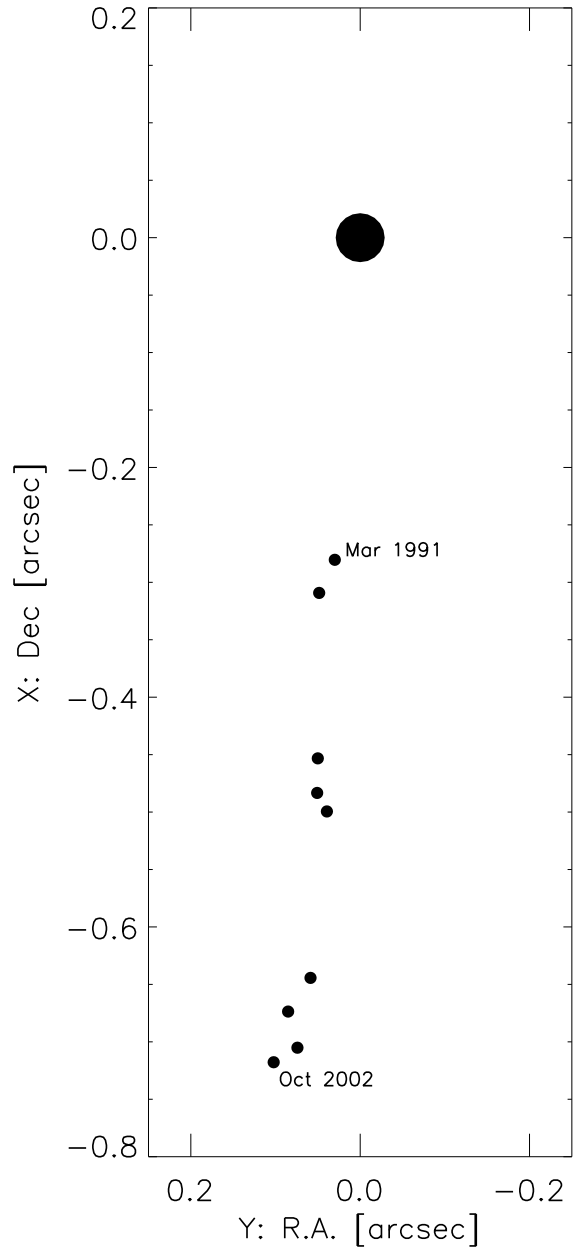}}

\caption{Orbital motion of the companion \object{EK Dra B} from 1991 to
2002. The primary is located at (0,0). As can be deduced from Table~1., the
last two points are separated by one year. Primary and secondary are thus
close to the apastron.  Combined with the RV measurements we estimate a period
of about $45\pm5$ years because Duquennoy \& Mayor observed the RV minimum in
May 1987.}
\label{orbital_motion} 
\end{figure}

\object{EK Dra} was repeatedly observed with the 3.5\,m-telescope on Calar
Alto from 1991 to 2002 using speckle interferometry, mostly in the K band
($\lambda = 2.2\,\mu\mathrm{m}$). An overview of these observations is given
in Table~\ref{speckle_obs} and is displayed in
Fig.~\ref{orbital_motion}. Observations 1 and 2 made use of a device for
one-dimensional speckle interferometry.  For these observations, the technique
of data acquisition and reduction has been described by Leinert \& Haas
(\cite{leinert89}). All other data points were obtained using the
near-infrared cameras MAGIC and OMEGA Cass that are capable of taking
sequences of short exposures (typically $\tau\approx$~0.1~s), and thus allow
speckle interferometry with two-dimensional detector arrays. Details of data
reduction and analysis have been described by K\"ohler et
al. (\cite{koehler00}). Briefly, $1000 - 1500$ short exposures (``frames'')
are taken for \object{EK Dra} and the nearby point source (PSF calibrator)
\object{BS~5436}. \object{BS~5436} is an F4IV star at a distance of
31.5~pc which implies that its relative size is $\leq 0.3$ mas, and thus it is
considerably smaller than the resolution of $1.22 \lambda/D = 160$\,mas of
the 3.5-m-telescope in the K-band. In Fourier space this reduces to
$\lambda/D=130$\,mas.  These images are stored in ``data cubes'' of 250
frames. The telescope position is switched between object and PSF calibrator
after each data cube to observe both under nearly identical atmospheric
conditions. After background subtraction, flat-fielding and correcting for bad
pixels, the data cubes are Fourier transformed. The modulus of the complex
visibility is derived by deconvolving the power spectrum of the object with
that of the PSF calibrator, while the phase is reconstructed using the
Knox-Thompson algorithm (Knox \& Thompson \cite{knox74}) and also the
bispectrum method (Lohmann et al. \cite{lohmann83}).  The complex visibility
is averaged over all observations of \object{EK Dra} taken on one
night. Finally, the binary parameters -- position angle, projected separation
and flux ratio $F_B/F_A$ -- are derived from a model fit to the complex
visibility in Fourier space. Except for the data points 1 and 2 in Table
\ref{speckle_obs}, for all observations the relative astrometry of the
components has been put into a consistent reference frame.  This reference
frame is primarily based on astrometric fits to images of the Orion Trapezium
cluster core, where precise astrometry has been given by McCaughrean \&
Stauffer (\cite{mccaughrean94}).

\subsection{Radial velocity measurements: data}

\begin{figure}[h]
\includegraphics[width=0.45\textwidth, angle=0]{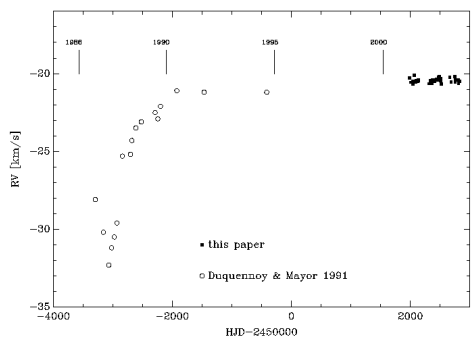}
\vspace{-3cm}
\caption{The RV data from Duquennoy \& Mayor (1991) marked with open circles
  and the three years of RV measurements obtained at the TLS marked
  with filled squares.}
\label{fig:rvall}
\end{figure} 

\object{EK Dra} is one of the stars monitored during the RV search
program for young and active stars of the Th\"uringer Landessternwarte
(TLS) described by Hatzes et al. (\cite{hatzes03}). For this program
we use the 2-m-Alfred Jensch telescope of the TLS, which is equipped
with an \'echelle spectrograph with a resolving power of
$\Delta\lambda/\lambda =67000$.  During the observations an iodine
absorption cell is placed in the optical light path in front of the
spectrograph slit. The resulting iodine absorption spectrum is then
superposed on the stellar spectrum providing a stable wavelength
reference against which the stellar RV are measured. In the first
step, the spectra are bias-subtracted, flat-fielded and extracted
using standard IRAF routines.
        
In the second step the RVs are calculated by modeling the observed
spectra with a high signal-to-noise ratio template of the star (without
iodine) and a scan of our iodine cell taken at very high resolution with
the Fourier Transform Spectrometer of the McMath-Pierce telescope at
Kitt Peak. The latter enables us to compute the relative velocity shift
between stellar and iodine absorption lines as well as to model the
temporal and spatial variations of the instrumental profile; see Valenti
et al. (\cite{valenti95}) and Butler et al. (\cite{butler96}) for a
description of the principles behind this technique.
Fig.~\ref{fig:rvall} shows our RV measurements together with those
obtained by Duquennoy \& Mayor (\cite{duquennoy91}).


RV measurements have been made at TLS since 2001 and these show that we can
achieve a routine RV precision of $\approx 3$\,m\,s$^{-1}$. However, our RV
measurements for EK Dra have an error of about 30\,m\,s$^{-1}$. Two factors
degrade the RV precision of our EK Dra measurements. First, EK Dra has a $v
\sin i$\ of $16.5\pm1.0$\,km/s. Since the RV error is proportional to the $v
\sin i$, the error compared to a more slowly rotating star with comparable
$S/N$ should be several times worse. Second, EK Dra is an active star and as
demonstrated by Saar \& Donahue (\cite{saar97}) the activity can introduce
significant RV ``jitter'' depending on the level of activity of up to several
tensof  m/s.

The relative RV measurements obtained by us were converted to absolute
values by measuring the absolute RV of the template spectrum by
fitting Voigt functions to photospheric lines with an equivalent width
larger than 0.1\,\AA \ in the wavelength range from 5000 to 6000\,\AA.

\subsection{Deriving the masses of the two components}

\begin{figure}[h]
\includegraphics[width=0.45\textwidth, angle=0]{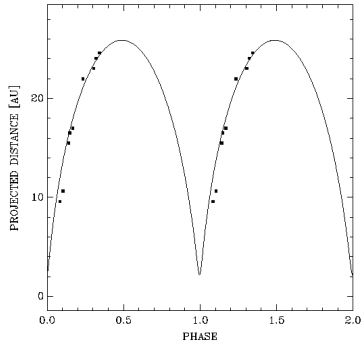}
\vspace{-3cm}
\caption{The projected distance versus time. The projected distance
was converted in to AU by using the {\sl Hipparcos} distance. The line
shows the fit using the elements given in Table\,\ref{elements}.}
\label{proje_dist}
\end{figure}

\begin{figure}[h]
\includegraphics[width=0.45\textwidth, angle=0]{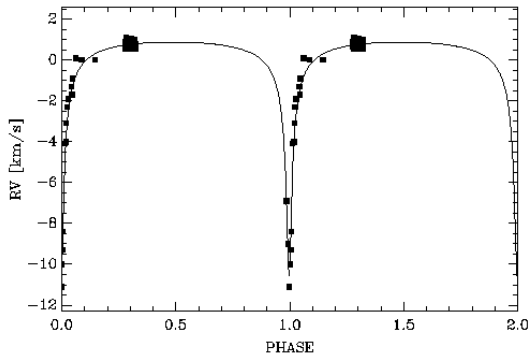}
\vspace{-3cm}
\caption{The radial-velocity curve together with the fit using the
values given in in Table\,\ref{elements}.}
\label{radial_velocity}
\end{figure}

\begin{figure}[h]
\includegraphics[width=0.45\textwidth, angle=0]{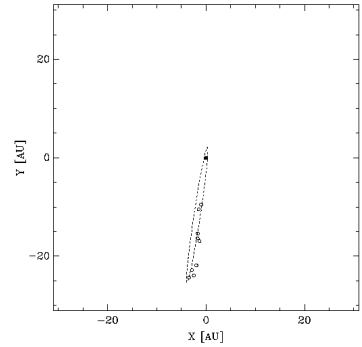}
\vspace{-3cm}
\caption{Using the distance as measured by {\sl Hipparcos} and the
radial velocity measurements it is possible to derive the true
masses of both components. Shown here is the derived orbit using the
values given in Table\,\ref{elements}.}
\label{orbit}
\end{figure}

The motion of \object{EK Dra B} with respect to the primary from 1991 to
2002 is shown in Fig.\,\ref{orbital_motion}. Although almost no
curvature is seen, this relative motion is definitely not caused by the
proper motion of \object{EK Dra A} with respect to a chance-projected
background star.  The proper motion of \object{EK Dra A} is
$\mu_{\alpha} = -138.6\,\mathrm{mas/yr}$ and $\mu_{\delta} =
-11.9\,\mathrm{mas/yr}$ ({\it Hipparcos} catalog), which is roughly
perpendicular to the observed motion and also much faster. Furthermore,
the companion has significantly slowed down over the time-span covered by
the observations, which can only be explained by orbital motion. The
slow-down additionally implies that the last observations are close to
the apastron. The spectroscopic observations of Duquennoy \& Mayor
(\cite{duquennoy91}) cover the periastron. By combining the radial
velocity data with the speckle observation it is possible to derive a
first orbit of the system and the true masses of both components using
the distance as measured by {\it Hipparcos}. The projected distance
versus time is shown in Fig.\,\ref{proje_dist}. The deceleration is clearly visible. By combining the speckle imaging data with the
radial-velocity data (Fig.\,\ref{radial_velocity}), we derive a period
of $45\pm5$~years. Given the orbital period the radial velocity
(Fig.\,\ref{radial_velocity}) constrains very well the mass-ratio of the
two components. By fitting an orbit to the speckle data
(Fig.\,\ref{orbit}) combined with the information on the projected
velocity and the {\it Hipparcos} distance, we can calculate the true
masses of the components and all other orbital elements by solving the
Keplerian equations (Kepler \cite{kepler1609}, Kepler
\cite{kepler1618}). The orbital elements are summarized in
Table~\ref{elements}. For the masses of the two components, we find
$0.9\pm0.1\,{\rm M}_\odot$ and $0.5\pm0.1\,{\rm M}_\odot$, for the
primary and the secondary, respectively. With $e=0.82\pm0.03$, the orbit is
surprisingly eccentric. The distance between the stars is nevertheless
still 2.2\,AU at the periastron. It thus seem unlikely that the
secondary has a big impact on the activity level of the primary.

%
%
\begin{table}
\caption{Orbital elements}
\begin{tabular}{ll}
\hline \hline
element & value \\
\hline
P           & $45\pm5$ yrs \\
$T_0$ [HJD] & $246960\pm30$ \\
e           & $0.82\pm0.03$ \\
$V_0$       & $-21.2\pm0.2$ km/s \\
$K_{1}$     & $5.8\pm0.2$ km/s \\
$K_{2}^{(1)}$ & $11.2$ km/s \\
$a^{(2)}$   & $14\pm0.5$ AU \\
$i$         & $85\pm3^\circ$ \\
$\Omega^{(3)}$ & $-98\pm1^\circ$ \\
$\omega^{(4)}$ & $180\pm10^\circ$\\
dist       & $33.94\pm0.72$\,pc \\
$f(m)={{M_2^3 \sin^3{i}}\over{(m_1+m_2)^2}}$ & $0.06\pm0.02 $ ${\rm M}_\odot$\\
$M_1$      & $0.9\pm0.1$ ${\rm M}_\odot$\\
$M_2$      & $0.5\pm0.1$ ${\rm M}_\odot$\\
\hline\hline
\end{tabular}
\\
$^{(1)}$  esitimated value using the mass-ratio of the primary
and secondary, and $K_1$\\
$^{(2)}$ semi-major-axis of the true orbit\\
$^{(3)}$ position angle of the line of intersection
     between the tangential plane of projection
     and the projected  orbital plane \\
$^{(4)}$ periastron longitude\\
\label{elements}
\end{table}

\section{Spectral synthesis analysis}

We observed \object{EK Dra} on September 11, 2001 and May 25, 2002
from Calar Alto using the high resolution \'echelle spectrograph FOCES
(Pfeiffer et al. \cite{pfeiffer98}) mounted on the 2.2\,m
telescope. Data reduction and analysis were carried out using the
reduction pipeline written in IDL especially for this fiber-coupled
spectrograph.

For the spectral synthesis analysis we used the model atmosphere code
MAFAGS. For a detailed description of the methods see Fuhrmann et
al. (\cite{fuhrmann97}). As described there, we deduce the effective
temperature from the Balmer line wings and the surface gravity from
the iron ionisation equilibrium and the wings of the Mg~Ib lines. The
analysis is performed strictly relative to the Sun. The method for
determining all stellar parameters was tested and compared extensively
in Fuhrmann (\cite{fuhrmann2004}).


\begin{table*}[ht]
\caption{Spectral parameters of \object{EK Dra} derived by spectral synthesis
  analysis.}
\begin{tabular}{lcccccc}
\hline\hline
name       & $T_{\rm eff}$ & $\log g$         & $[Fe/H]$         & $v \sin i$
& M$_V$ & M$_{\rm bol}$ \\
\hline
EK Dra     & $5700\pm70$   & $4.37\pm0.10$ & $-0.16\pm0.07$ & $16.50\pm1.00$ & 7.60 &  4.79 \\ 
\hline
\hline
\end{tabular}
\end{table*} 

\subsection{Results for \object{EK Dra}}

\begin{figure*}[!ht]
\includegraphics[width=0.8\textwidth, angle=0]{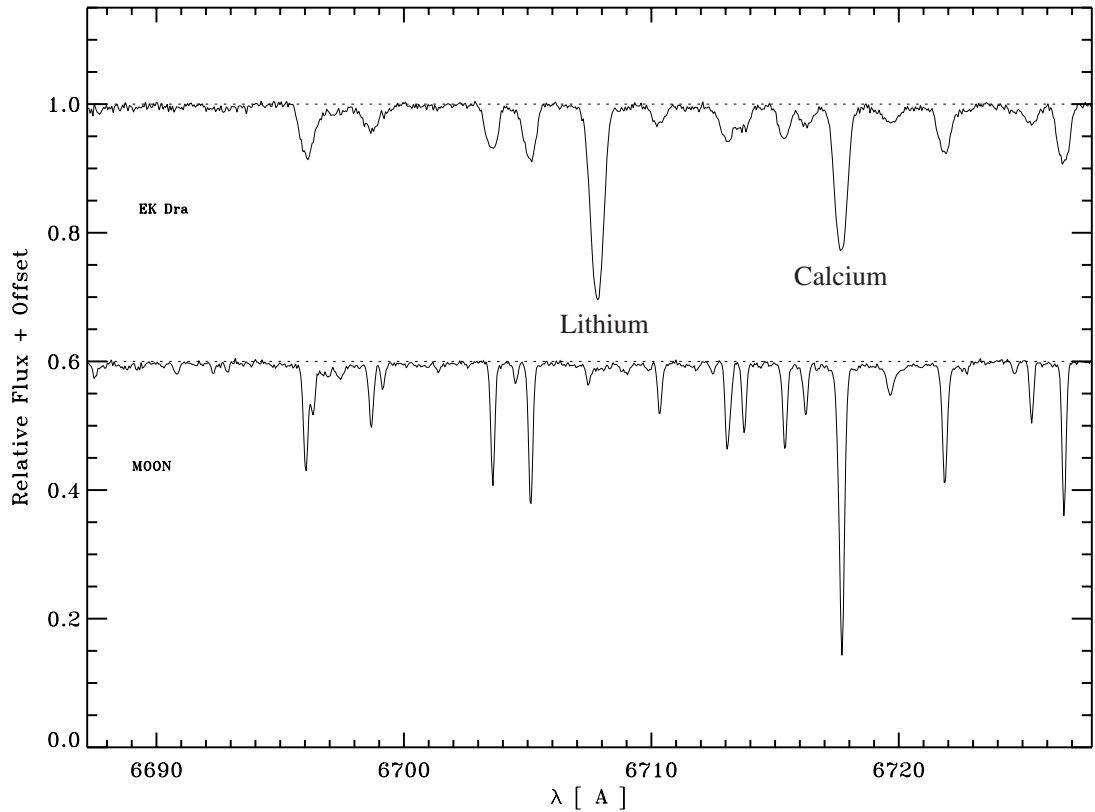}
\caption{The lithium absorption line at 6707.8\,\AA~in the spectrum of
  \object{EK Dra} compared to reflected sunlight on the moon. Note the
  high $v \sin i$ of $16.5\pm1.0$\,km/s of \object{EK Dra}. The strong
  lithium line indicates that EK~Dra is young.}
\label{fig:li}
\end{figure*}

The chromospheric activity, the variability, the presence of core filling-in
of the H$\alpha$, the calcium H \& K and the magnesium Ib-lines and a strong
lithium absorption line at 6707\,\AA~(Fig.~\ref{fig:li}) indicate that the
star is indeed young and if we believe the star to belong to the Pleiades, we
can assume an age of about 125\,Myr.

\begin{figure*}[ht!]
\includegraphics[width=0.8\textwidth, angle=0]{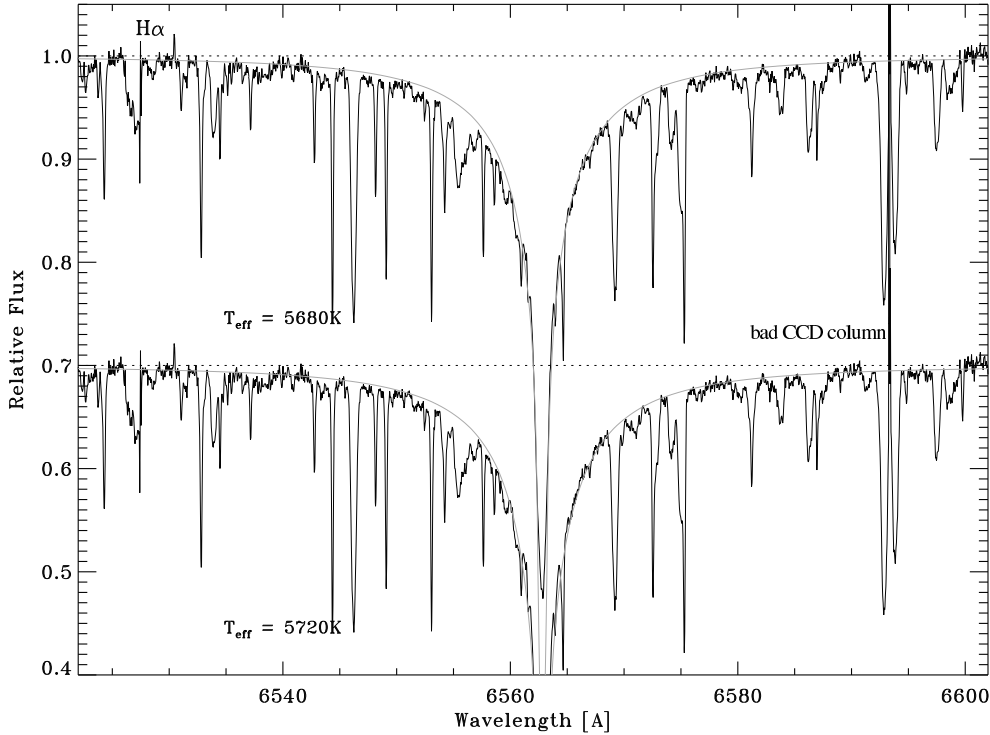}
\vspace{-5cm}
\caption{The H$\alpha$-line together with two theoretical H$\alpha$-line
  profiles to estimate the temperature. The real temperature is in between the
  plotted ones. Note the filling in of the H$\alpha$-line.}
\label{fig:teff}
\end{figure*} 

The spectral syntheses analysis we performed was challenging because
it is known that the star rotates fast ($16.5\pm1$\,km/s) and it
exhibits huge stellar spots which are much cooler that the surrounding
surface. Depending on the position of the star spot, the star could
appear cooler than it actually is. On the other hand the
H$\alpha$-line core is filled-in up to a level of 0.5. The
H$\beta$-line is not noticeable filled-in. To measure the effective
temperature we calculate a grid of line profiles of the H$\alpha$- and
H$\beta$-lines using the surface gravity, and iron abundance
determined using the Fe I \& II- and Mg Ib-lines. We fit the wings of
the strong lines but not the core. The final effective temperature is
measured using the H$\alpha$-temperature and comparing it with the
H$\beta$-temperature giving a weight of 75\% to the
H$\beta$-temperature. If the new derived temperature changed more than
50\,K compared to the previously obtained temperature we repeat
the determination of the iron abundance as well as the surface
gravity.

In the case of the spectrum taken on Sep. 11, 2001, both temperatures
were identical while in the spectrum taken on May 25, 2002 the
H$\beta$-line leads to a 40\,K hotter star. For a fit to the
H$\alpha$-line profile see Fig.~\ref{fig:teff}. However, the
temperature most consistent with all measurements is
$5700\pm70$\,K. We have double-checked the derived effective
temperature with a fit to the H$\alpha$ and H$\beta$ lines, assuming a
spot temperature of 4500\,K and a spot coverage of 1/4 of the visible
surface. The resulting measured temperature does not significantly
change the previously obtained results of the one-temperature fit
because the continuum of the spot is only 10\% of the continuum level
of the surrounding stellar surface. This is because the spot is much
cooler than the photosphere, and hence its total contribution to the
light emitted from the star is very small.  This results in a
correspondingly small contribution to the total spectrum. In fact, the
spot only produces a little hump in the photospheric line-profiles
used for the Doppler imaging, which changes the EW of a photospheric
line only by a small amount, and cannot be seen in H$\alpha$ and
H$\beta$. Thus, even for a highly spotted star, the average line
spectrum is dominated by the photosphere.

Comparing the spectroscopic distance determined by us to the Hipparcos
parallax, we have a 5.8\% discrepancy. This makes us confident that
the spectral parameters we derived, especially the surface gravity
$\log g$, are reasonable. Regarding the H$\alpha$ line depth and the
line wings (Fig.~\ref{halpha}, lower panel), we see significant
changes of the effective temperature from 5700\,K to 5580\,K when
using only the H$\alpha$~temperature. The H$\beta$~temperature is not that
strongly affected.

The analysis from Wyse \& Gilmore (\cite{wyse95}) lead to a somewhat
lower iron abundance of $-0.214$\ which was estimated by narrow-band
Str\"omgen photometry. The temperature of \object{EK Dra} of 5930\,K
estimated by Dorren \& Guinan (\cite{dorren94}) is higher than our
measured temperature of $5700\pm70$\,K. But in our case the H$\alpha$-
and H$\beta$-line profiles would not support such a high temperature
in our spectrum. Eggen (\cite{eggen98}) also estimated a metallicity
of \object{EK Dra} of $-0.24$\,dex using Str\"omgen photometry, which is
lower than the metallicity measured by us of $-0.16$\,dex.

For their Doppler imaging, Strassmeier \& Rice (\cite{strassmeier98})
need stellar parameters as an input and they used Kurucz model
atmospheres where they fixed the iron abundance to $[Fe/H] = 0.0$ and
the surface gravity to $\log g =4.5$\ to derive $T_{\rm eff} =
5870\pm50$\,K and $v \sin i=17.3$\,km/s. They recover stellar spots at
high latitude with a temperature difference of $\Delta T = 400$\,K. A
polar spot cannot be confirmed or excluded by their data. The
inclination of the stellar rotation axis is $\sim 60^\circ$. The
inclination of the orbit thus is significantly different from $\sim
60^\circ$. Because of the significant difference from our spectral
parameters (especially the effective temperature and the iron
abundance) we propose caution when using the conclusions of the
Strassmeier \& Rice (\cite{strassmeier98}) Doppler imaging.

\section{Space motion and age of EK~Dra}
\label{UVW}  
\cite{soderblom1987} estimate the age of EK~Dra to be $\sim
70$\,Myr using activity indicators and claim that the star could have
traveled from the Pleiades to the solar vicinity with a peculiar
velocity of only 2\,km/s. On the other hand, based on activity
indicators \cite{wichmann2003a} claim that the star is even younger
than the Pleiades with an age of $\sim 50$\,Myr and call it a member
of the local association. \cite{wichmann2003b} have traced the space
motion backward in time and exclude a former membership of the young
associations Lupus-Centaurus-Crux or Upper-Centaurus-Lupus.

Stauffer et al. (\cite{stauffer98}) derived an age of the Pleiades of
about 125\,Myr using the aproproate distance scale and modern stellar
evolution calculations. Estimates by \cite{basri1996} using two
brown dwarf members of the Pleiades give an age of 115 to 125\,Myr.

We have used the proper motion $\mu_\alpha=-138.61\pm0.72$\,mas,
$\mu_\delta=-11.92\pm0.56$\,mas and the parallax $\pi=29.46\pm0.61$
measured by {\it Hipparcos}, as well as the RV of
$-23.1\pm0.2$\,km/s measured by us to calculate the galactic space
motion $(U/V/W)=(1.3\pm0.3/24.8\pm0.4/-0.3\pm0.3)$\,km/s).

EK~Dra is located in the vicinity of the Sun at a distance
measured by {\it Hipparcos} of $33.94\pm0.72$\,pc. \object{EK Dra} is
a young object and it is likely that its age lies within 50 to
125\,Myrs depending on the criteria one applies, e.g. if one only
regards the activity (50\,Myr) or if one assumes it is a Pleiades
field star (125\,Myr).

With the average absolute brightness of the primary of
$M_K=3.4\pm0.1$\,mag, and of $M_K=6.0\pm0.2$\,mag for the secondary
component, and using the evolutionary tracks published by Baraffe et
al. (\cite{baraffe98}) (Model: [M/H]=0, $\alpha=1.9$, Y=0.282), one
derives a mass of $1.025\pm0.100$\,M$_\odot$\ and an age of 35\,Myr
(lower limit: 30\,Myr and upper limit 1.5\,Gyr on the main
sequence (MS)) for the primary. The mass derived from the
evolutionary tracks agrees with the true mass within the errors. The
companion is expected to be about 6\,mag fainter than the primary in
the V-band. The visible spectrum of \object{EK Dra} is thus completely
dominated by the primary.

\section{Activity, line asymmetry and RV variations}

\begin{figure}[h]
\includegraphics[width=0.45\textwidth, angle=0]{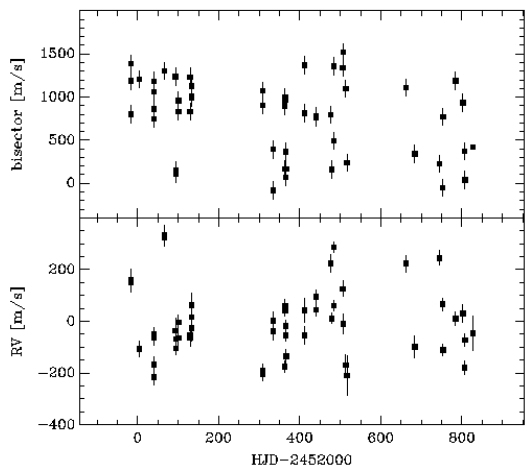}
\vspace{-2cm}
\caption{Top panel the bisector analysis of the chromospheric inactive
  lines outside the region which is affected by the iodine
  lines. Lower panel the original RV data observed in the years 2001,
  2002 and 2003 at the TLS.}
\label{rv_data}
\end{figure}

\begin{figure}[h]
\includegraphics[width=0.45\textwidth, angle=0]{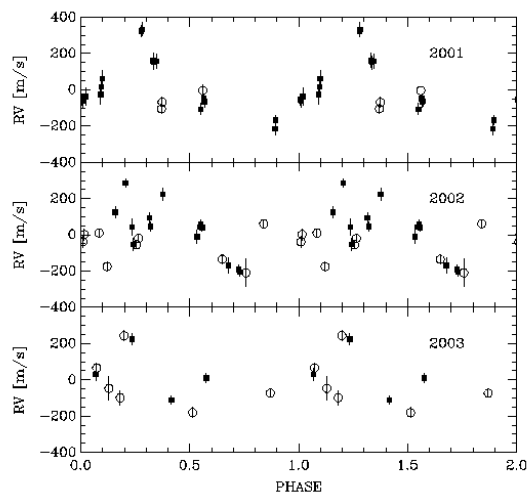}
\vspace{-2cm}
\caption{The RV data phase folded with the period of 2.769 days and split
  between the different years of observation. The filled symbols represent 
  the asymmetric lines and the open circles the symmetric lines.}
\label{rv_years}
\end{figure}

\begin{figure}[h]
\includegraphics[width=0.45\textwidth, angle=0]{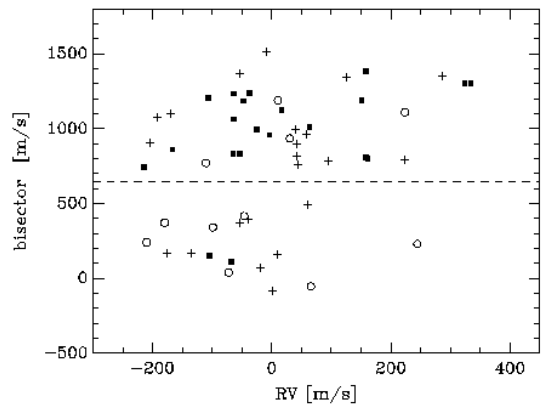}
 \vspace{-2cm}
\caption{The results of the bisector analysis for all three years of
observations. The filled squares are the data from 2001, the crosses
the data from 2002, and the circles from 2003. The spectra from 2001
mostly have asymmetric lines, whereas in 2002 we observe asymmetric as
well as symmetric lines. In 2003 the lines are mostly symmetric. The
dashed line indicates the border between the symmetric and
asymmetric spectral lines. }
\label{assym}
\end{figure}

\begin{figure}[h]
\includegraphics[width=0.45\textwidth, angle=0]{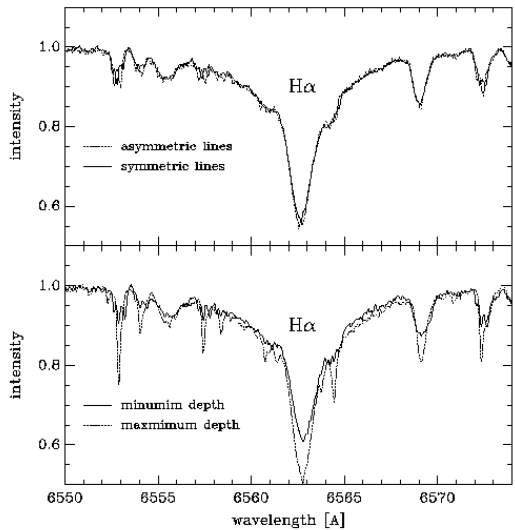}
\vspace{-2cm}
\caption{The upper panel shows the average profile of H$\alpha$ of all
spectra where the photospheric lines are symmetric in comparison to the
average profile of all spectra where the photospheric lines are
asymmetric. The difference between the two H$\alpha$ profiles is
remarkably small. The only notable difference is that the core of
H$\alpha$ is a little deeper in the case of the spectra where the
photospheric lines are asymmetric. 
The lower panel shows that there are significant variations in the 
depth of H$\alpha$. Also the wings of the H$\alpha$ lines show a difference
and lead to a 120\,K cooler star. This effect is smaller in H$\beta$.}
\label{halpha}
\end{figure} 

Because EK\,Dra\,B was close to the apastron at the time of the
RV observations, and since the period of this highly eccentric orbit
is $45\pm5$ years, the influence of the secondary on this section of
the RV curve is negligible (Fig.\,\ref{radial_velocity}). With an
accuracy of our relative RV measurements of 30\,m/s it is possible to
study the RV variations caused by stellar activity. Saar \& Donahue
(\cite{saar97}) have studied the activity-induced RV variations and
found a relation of $A_{S} \approx 6.5 f^{0.9}_{S} v \sin i$, where
$f_{S}$ is the star-spot area coverage [$f_{S}$ is in \%, $ v \sin i$
in km/s, and $A_{S}$ in m/s]. For spots located at the equator, the
filling factor is given by this equation. For spots located close to
the pole, the RV variations would give a rise to a too small filling
factor. Because spots on the pole do not give rise to RV variations,
the amplitude of the RV variations in principle just give a lower
limit for the filling factor $f_S$. From the measured RV variations we
thus derive (Fig.\,\ref{rv_data}, upper panel) the peak-to-peak
amplitude variation of about 550\,m/s, which implies that $f_{S}\geq
5.2\%$ due to the fact that we suspect the presence of a polar spot.

Performing a period analysis we find a period of the RV variation of
$2.767\pm0.005$\,days in the first two years of observations but
adding the third year the period search programs do not recover the
period clearly. When we inspect the data (see Fig.~\ref{rv_years}), we
see that the amplitude of the RV signal declines.

Moreover, we see a change in the bisector of the spectral lines over
the three years of observations. We use the classical Doppler imaging
lines, Fe\,I at 6400\,\AA, and Ca\,I at 6439\,\AA\ to measure the
amplitude of the asymmetry because they are not blended by other
weaker spectral lines and there are no iodine lines in this part of
the spectrum.  Shown in Fig.\,\ref{rv_data} (upper panel) is the
average velocity difference between the lower and the upper half of
the spectral lines.  A negative sign means that the lower part is
blue-shifted with respect to the upper part of the line.  During the
three years the behavior of the bisector significantly changed. In
2001, the velocity difference is relatively constant with a shift of
700 to 1500\,m/s, whereas in 2003, the bisector amplitude varies
between 1200\,m/s and $-100$\,m/s. We then can split the lines between
asymmetric lines with a bisector from 700\,m/s to 1500\,m/s and
symmetric lines with a bisector of 600\,m/s to $-500$\,m/s
(Fig.~\ref{assym}). Excluding the symmetric lines from the period
analysis we recover the period over all three years.  Compared to the
amplitude of the RV signal of the symmetric lines, the signal for the
asymmetric lines is more pronounced (Fig.~\ref{rv_years}).

We can summarize our finding as follows: in 2001 and 2002, we observe
a periodic RV signal and the spectral lines were quite
asymmetric. This RV signal is less obvious in the data from 2003 but
can be recovered if only the asymmetric lines are used. Clearly, this
is the signal of stellar spots not located at the pole; the Doppler
imaging by Strassmeier \& Rice (\cite{strassmeier98}) shows the
presence of spots close to the equator. One possible explanation would
be that in 2003, there were fewer spots at mid latitudes, or
alternatively, there were so many plage regions so that it is
difficult to recover the periodic RV signal from the spots. Plage
regions are known to produce less asymmetric lines.

In order to find out if changes in the symmetry of the lines are
related to the activity level, we averaged all spectra with symmetric
photospheric lines and all spectra with asymmetric photospheric
lines. Fig.\,\ref{halpha} shows the H$\alpha$-line profiles of all
spectra where the photospheric lines are symmetric in comparison with
the average profile of all spectra where the photospheric lines are
asymmetric. The difference between the two H$\alpha$ profiles is
remarkably small. However, we do see significant variations of the
depth of H$\alpha$. Thus testing if the amplitude of the
RV signal is dependent on the depth of H$\alpha$, we calculate the
average RV of all spectra where H$\alpha$ is deepest and all spectra
where H$\alpha$ is shallowest. We do not see any significant
difference between the two samples, as the average RV of the first
sample is $-40\pm80$~m/s$^0$\footnotetext{variance}, and the average
of the second $-130\pm120$~m/s\,$^0$. There is also no significant
difference in the asymmetry of the two, as the average asymmetry of
the first sample is $860\pm380$~m/s\,$^0$\ and the asymmetry of the
second sample is $1000\pm350$~m/s\,$^0$. Thus, there is no obvious
correlation between the RV or the asymmetry and the depth of the
H$\alpha$-line. Additionally, there is only a very weak correlation
between the asymmetry and the RV.

The change of the temperature as seen in Fig.~\ref{halpha} (lower panel)
of about 120\,K measured by the H$\alpha$-line profile is not as visible in
the H$\beta$-line profile. There, it only shows a temperature change of 20\,K
which lies within the errors of this method. Giving more weight (75\%) to the
H$\beta$ temperature will result only in a small temperature change. But of
course the temperature change will also affect the Fe I and Fe II ionisation
equilibrium and this will have an effect on the determination of the surface
gravity $\log g$. We cannot verify this here because the spectra obtained at
TLS are mostly observed with the iodine cell in the light path which
superposes iodine absorption lines on the part of the spectrum containing the
necessary iron lines.

How can we interpret these findings? A spot that is located close to the
equator would lead by its appearance to a positive asymmetry and also to
a redshift of the spectral line.  When the spot is receding due to
rotation, it would lead to a negative asymmetry and to a blue-shift of the
line. In this case we would expect a clear correlation between the
asymmetry and the RV. Such a correlation is for example observed in RS
CVn systems (Donati et al. \cite{donati95}).  This is clearly not what
we observe. The Doppler imaging of this star mainly shows a polar spot,
and our estimate of the spot size from the amplitude of the RV
variations indicates that there are only a few spots close to the equator.

In the case of sunspots it is well known that the Evershed effect
causes the line cores to be blue-shifted on the limb-side of the
penumbra and redshifted at the other side. The blue-shifted lines show
a blue asymmetry that is negative, whereas the red-shifted
lines show a positive asymmetry (Sanchez Almeida et al.
\cite{Sanchez96}). While the contribution of the total light emitted
is quite small, a spot does produce a hump in the profile of the
photospheric line. Because we can measure the RV to a very high
accuracy, such effects can be detected. However, observations of the
penumbra of spots close to the disk center show positive as well as
negative asymmetry (Balasubramaniam \cite{bala02}). Thus it is not
surprising that in the case of a star where there are numerous spots
close to the pole, there is no correlation between the RV and
the asymmetry. It is interesting to note that the bisector of the
solar granulation shows the famous C-shape that will result in
almost no shift between the upper half of the line and the lower half.
We would call such lines symmetric. However, once the convective
structures are spatially resolved, the red-shifted inter-granular
lanes turn out to give rise to spectral lines with a negative
asymmetry, and the granules give rise to lines with a slightly
positive asymmetry (Guenther \& Mattig \cite{guenther91}).

Thus we presume that large and complicated flow patterns in the polar
spots prevent us from observing a correlation between the RV and
line asymmetry or depth of H$\alpha$. Nevertheless our observations show that
the periodic RV signal can be detected if only the lines with positive
asymmetry $\geq 600$\,m/s are taken into account. This indicates that there is
a link between the asymmetry and RV, which implies that both are possibly
caused by spots.

Our observations also suggest that in the case of \object{EK\,Dra} it is
impossible to account for the RV signal of the star spots to increase
the sensitivity of a possible RV signal of an orbiting planet because the
correlation of either the depth of H$\alpha$\ or the line-asymmetry with the
rotation phase is limited.


\section{Summary and conclusions}

We have studied the young star EK\,Dra spectroscopically and have
resolved the binary system by means of speckle interferometry. This
star is in fact a young star evolving analogously to the Sun, where
the analogy in the evolution is even closer than previously thought.

\begin{itemize}
\item By combining the speckle interferometry with old and new RV
      measurements we have for the first time a reliable mass estimate
      of EK\,Dra\,A and EK\,Dra\,B of $0.9\pm0.1$\,M$_\odot$\ and
      $0.5\pm0.1$\,M$_\odot$.
\item The orbit of the binary turns out to be highly eccentric with
      $e=0.82\pm0.03$. The period of the system is of $45\pm5$\ years, 
      much longer than previously thought.
\item The high resolution spectra allow us to derive a new $T_{\rm
      eff}$ of $5700\pm70$~K and a new value of $\log g$ of
      $4.37\pm0.10$. Additionally, we measured a metallicity of
      $[Fe/H]=-0.16\pm0.07$.  The new values thus differ from those of
      Strassmeier \& Rice (1998), who found $T_{\rm
      eff}=5870\pm50$\,K, $\log g = 4.5$\ and the metallicity
      $[Fe/H]=0.0$.  However, the spectroscopic distance derived using
      our new values now is in good agreement with the {\it Hipparcos}
      distance.
\item The Baraffe et al.~(\cite{baraffe98}) models allow a mass
      determination ($1.025\pm0.100$\,M$_{\odot}$) and an age
      determination of 35\,Myr (30\,Myr to MS) for the primary.
\item From our data we can exclude the presence of a third low-mass
      companion in the system as suggested by \cite{stanimir2004}.
\item We also see variations in the H$\alpha$~line wings which lead to a
      temperature difference between the two spectra of 120\,K. This
      temperature change could be caused by huge stellar spots. The effect on
      the line core and wings of the H$\beta$~lines is not as strong.
\item The precise RV measurements taken in the course of the program
      shed more light on the activity of the star. We find that the
      primary star shows large amplitude RV variations which are
      presumably caused by its high activity level. In 2001 and 2002
      the signal was periodic with a period of $2.767\pm0.005$\,days.
      However, it seems that this period is slightly different from
      the photometric rotation period (2.599 to 2.796, Strassmeier \&
      Rice \cite{strassmeier98}). This difference could be caused by
      differential rotation of the star. In 2002 this signal is
      significantly less obvious, and in 2003 almost absent. The
      periodic signal can be recovered from the whole dataset if only
      spectra in which the photospheric lines are asymmetric are
      used. We interpret this signal as being due to spots that are
      not located at the pole. However, the correlation between
      asymmetry and RV is weak. If the same analysis is performed on
      spectra in which the photospheric lines are nearly symmetric, we
      cannot recover the period signal of the RV variations. Also
      there is no obvious correlation between H$\alpha$ and either the
      RV or the asymmetry. We interpret this by assuming that the star
      has large spots on the pole and possibly the Evershed-flow of
      these spots affects the RV, as well as the asymmetry signal.  We
      conclude that in the case of \object{EK Dra} the RV signal is
      related to the rotation period of the star and is caused by its
      activity cannot be simply subtracted from the RV signal to
      increase the sensibility for the detection of a possible third
      object.
\end{itemize}

\begin{acknowledgements}
This research has made use of the SIMBAD database, operated at CDS,
Strasbourg, France. The authors want to thank Klaus
Fuhrmann for the useful discussion. J.W. acknowledges support from the
Deutsches Zentrum f\"ur Luft- und Raumfahrt under grant number 50 OR
0009. The data reduction made use of the ``Binary/Speckle'' software
package developed by Rainer K\"ohler.
\end{acknowledgements}

\end{document}